\documentclass[aps,unsortedaddress,frontmatterverbose,print,10pt, twocolumn]{revtex4-2}

\usepackage{graphicx}% Include figure files
\usepackage{dcolumn}% Align table columns on decimal point
\usepackage{bm}% bold math
\usepackage{xcolor}
\usepackage{amsmath}
\usepackage{booktabs}
\usepackage{textcomp}
\usepackage{lipsum}
\usepackage{amssymb}
\usepackage{booktabs}
\usepackage{easy-todo} %does what it says on the can.
\usepackage{xr} %used to ref SI figures/tables
\begin{document}
    \preprint{APS/123-QED}

    \title{Method for real-time monitoring of paramagnetic reactions using spin relaxometry with fluorescent nanodiamonds}% Force line breaks with \\

    \author{Trent Ralph}
    \email{trent.ralph@unimelb.edu.au}
    \affiliation{School of Physics, The University of Melbourne}%
    \affiliation{Manufacturing, CSIRO}%
        
    \author{Erin S. Grant}
    \email{erin.grant@rmit.au}
    \affiliation{School of Science, RMIT University}%

    \author{Lianne Lay}
    \affiliation{School of Physics, The University of Melbourne}%
    
    \author{Sepehr Ahmadi}
    \email{sepehr.ahmadi@csiro.au}
    \affiliation{Manufacturing, CSIRO}%

    \author{David A. Simpson}%
    \email{simd@unimelb.edu.au}
    \affiliation{School of Physics, The University of Melbourne}%

\begin{abstract}
Spin relaxometry using fluorescent nanodiamonds (FNDs) has been applied successfully to sense numerous paramagnetic target molecules such as free radicals and metalloproteins. However, despite their high sensitivity, $T_1$ spin relaxation measurements are often hampered by their slow acquisition speed. Here, we demonstrate a method that allows for real-time monitoring of paramagnetic chemical reactions. We demonstrate $T_1$ spin relaxometry from thousands of FNDs using an optimised cuvette-based system integrating an avalanche photodiode operated in linear mode, and a fast, field-programmable gate array (FPGA) for data collation. We demonstrate chemical monitoring of the reduction of Cu(II) to Cu(I) ions in-solution with a 15 second integration using an optimised $T_1$ sensing protocol. Our method achieves more than two orders of magnitude speed up with an order of magnitude reduction in cost when compared with traditional techniques. With further technical improvements, we believe this in-solution method could be extended to sense the sub-second chemical kinetics of paramagnetic molecules in solution.
\end{abstract}

\maketitle

\section{Introduction}

The nitrogen-vacancy (NV) defect in diamond has numerous sensing applications, from nanoscale electron paramagnetic resonance (EPR) of a single protein~\cite{shiSingleproteinSpinResonance2015}, to \textit{in situ} temperature sensing via optically detected magnetic resonance (ODMR) ~\cite{sotomaDiamondNanothermometry2018, choePreciseTemperatureSensing2018}. 
Many NV sensing protocols monitor changes in the spin coherence ($T_2$) or spin relaxation ($T_1$) times of the defect to infer changes in the local magnetic field ~\cite{mzykRelaxometryNitrogenVacancy2022,hongNanoscaleMagnetometryNV2013}. $T_1$ spin relaxometry is one of the more popular techniques used to probe local fluctuating magnetic fields. 
Detection of paramagnetic species in liquid samples can be achieved in various geometries, using both single crystal diamond ~\cite{simpsonElectronParamagneticResonance2017} or fluorescent nanodiamond (FND) ensembles . In this work we focus on the latter as they provide cost and sensitivity advantages over single crystal diamond\cite{gorriniFastSensitiveDetection2019}.

The spin relaxation time $T_1$ from NV containing FNDs have been measured using two main approaches; confocal microscopy is often used for single FNDs studies that require high spatial resolution and employ sensitive single photon counting avalanche photodiodes (SPADs) and time-correlated single-photon counting devices. 
Alternatively, widefield fluorescent microscopes can be deployed to measure FND ensembles by utilising sensitive scientific CMOS cameras or amplified photodiodes. 
In a previous report, we presented a third approach using an optical cuvette which allowed measurement of the $T_1$ spin relaxation time from FND ensembles in solution. However, the use of a SPADs detector and time correlated single photon counter limited the acquisition times to $\sim$50 minutes making dynamic spin relaxation measurements impractical.   

Here, we demonstrate enhanced in-solution $T_1$ sensing with a two orders of magnitude improvement in measurement speed and an order of magnitude reduction in cost. 
Using this method, we are able to perform real-time monitoring of a paramagnetic chemical reaction with a time resolution of 15 seconds using a three point $T_1$ acquisition sequence. 
The realisation of a real-time, low cost system for the monitoring of chemical reactions, brings $T_1$ spin relaxometry using FNDs into a regime where it can be applied ubiquitously across the interdisciplinary fields of physics, chemistry and biology.

\section{$T_1$ spin relaxometry using fluorescent nanodiamonds}
\begin{figure*}
    \centering
    \includegraphics[]{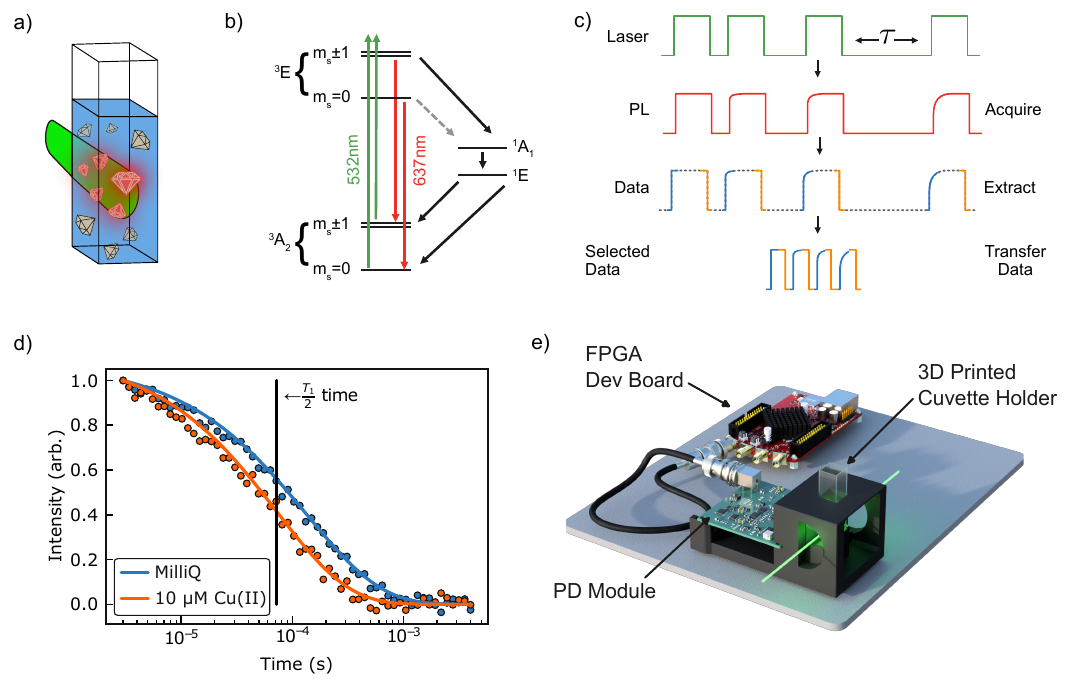}
    \caption{a) In-solution measurements are conducted on dispersed FNDs in a standard cuvette using a green (532~nm) excitation laser which generates red fluorescence from the NV defects b) Schematic of the energy levels of the NV defect consisting of ground($^3A_2$) and excited state triplets($^3E$) and two intermediate singlet states ($^1A_1$,$^1E$).After excitation with green (532~nm) the system can decay back to the ground state radiatively (637~nm) or non-radiatively through the intermediate singlet states. c) A schematic of a the laser pulse sequence, resulting fluorescence, and transmitted data. d) A typical $T_1$ relaxation curve resulting from the processed data acquired from the pulse sequence in c. Plot compares the baseline measurement of FND dispersed in MilliQ (blue trace), and the same sample after addition Cu(II) ions (10~\textmu M) (orange trace). e) Rendering of the 3D printed cuvette holder with integrated PD module connected to FPGA development board which acquires the voltage signal from the PD module.}
    \label{fig:1}
\end{figure*}
In-solution spin relaxometry utilises FNDs suspended in a conventional cuvette, as shown Figure \ref{fig:1}a, allows for streamlined sample preparation and interrogation of millions of FND simultaneously.
$T_1$ relaxometry takes advantage of the electronic structure of the NV defect shown in Figure \ref{fig:1}b.
Under 532~nm excitation, the negatively charged NV defects within the FNDs emit broadband red fluorescence with a zero phonon line at 637 nm. When excited with a sufficiently long (10s of~\textmu s laser pulse) the NV defects can be spin-polarised into the $m_s=0$ ground state. The NV fluorescence is spin-dependent with the $m_s=0$ brighter than the $m_s=\pm{1}$ ground spin state. Together, these attributes provide a method for measuring the spin lifetime $T_1$ of the NV $m_s=0$ ground state, using the optical pulse sequence shown in the inset of Fig \ref{fig:1}c.  
This all-optical protocol, previously reported \cite{grantMethodInSolutionHighThroughput2023}, consists of laser pulses interspersed by variable evolution-times, $\tau$. Each laser pulse is used to both read-out the spin-state of the NV ensemble and then repolarise the system before the next evolution period.
During each evolution time, the system interacts with the surrounding environment wherein local magnetic fluctuations cause enhanced relaxation of the polarised state back to a thermal equilibrium. 
By varying the $\tau$ time, a $T_1$ relaxation curve of the kind shown in Fig \ref{fig:1}d, can be generated.

The baseline value of $T_1$, determined by phononic interactions and spin-noise intrinsic to the FND, is decreased by external magnetic noise that overlaps with the zero-field splitting (2.87~GHz) of the NV defects. 
Measurements of the change in spin relaxation time away from the baseline, can therefore be used to sense magnetic noise sources such as paramagnetic ions, metalloproteins, or free radical species. 

%%%%%%%%%%%%%%%%%%%%%%%%%%%%%%%%%%%%%%%%%%%%%%%%%%%%%%%%%%%%%%%%%
\section{Implementation method}
The $T_1$ spin relaxometry method presented here utilises an avalanche photodiode (PD) (C10508-01, Hamamatsu) operating in linear mode, coupled with a low cost, fast, field-programmable gate array (FPGA) based data acquisition board (STEMLab 125-14, Red Pitaya)\cite{KumarHighDynamicRange2024, HuMultiPoint2023, hesseDirectControlHigh2021, wangHighPerformanceDigitalLockIn2023, neuhausPythonRedPitaya2024}. 

The PD-based system consists of a 532~nm (GEM, Laser Quantum) excitation laser, modulated with an acousto-optical modulator (AOM)(R35085-5, Gooch \& Housego).
Laser pulses are generated by gating the AOM using a pulse blaster (PB24-100-4k-USB-RM, SpinCore).
The excitation beam (300~mW) is focused into the center of the cuvette with a 150~mm bi-convex lens (LB1437-A-ML, Thorlabs). 
The signal collection consists of a custom, 3D printed cuvette holder with integrated band-pass filter (BrightLine 731/137, Thorlabs) and photodiode (PD) module, shown in figure \ref{fig:1}e.
The use of a custom, 3D printed cuvette holder was required to limit the stand-off between the sample and PD to maximize the NV collection efficiency. 
The 3D printed holder also incorporates an 4~mm aperture to limit high-angle scattered excitation light which can pass through the band-pass filter, reducing signal contrast.  

The Red Pitaya FPGA is used to acquire the voltage signal generated by the PD module.
The Red Pitaya is an open-source development board based around the Xilinx Zynq system on a chip (SoC).
It contains both FPGA based programmable logic and an accompanying Arm based processing system. 
The board is able to sample the voltage signal at up to 125~MS/s, which far exceeds the bandwidth of the PD module (10~MHz). 
The programmable logic is used to process the signal in real time by filtering, followed by decimation/averaging before writing the data to on-board RAM. 
In order to minimise data transfer times, the signal was sampled at the full 125~MS/s, however, every 4 samples were averaged into a single sample resulting in an effective sampling rate of 31.25~MS/s. This reduced data transfer requirements while maintaining a Nyquist frequency above the bandwidth of the PD module.
To minimise any aliasing from the decimation process, a digital decimation filter was also applied using the programmable logic.

Each laser pulse is applied for 100~\textmu s to archive a maximally polarised state --- evidenced by a plateau in fluorescence that occurs in the final 10~\textmu s. 
The optimal spin contrast is achieved by integrating the NV photoluminescence (PL) signal over the first 10~\textmu s of each laser pulse.\cite{grantMethodInSolutionHighThroughput2023} 
Each data point can be normalised by dividing the NV PL signal collected from the first 10~\textmu s of the excitation pulse, with the NV PL signal from the final 10~\textmu s of the pulse. 
This nomalisation removes PL variations due to laser intensity fluctuations and/or slow modulations arising from NV charge state changes. 
As a result, the PD signal during the middle (80~\textmu s) of each pulse, as well as the dark evolution time between each pulse, is not used and can be excluded to improve the data transfer efficiency. This selected data transfer resulted in an order of magnitude speed-up in the overall data acquisition. 

The acquired data is fitted to a stretched exponential of the form:
\begin{equation}
    I(\tau) = \mathcal{C}\exp((\tau/T_1)^p)+I(\infty),
    \label{eq:fit}
\end{equation}
where $\mathcal{C}$ is the spin contrast, and $p$ is a stretch parameter that accounts for the distribution of $T_1$ relaxation times that contribute to the overall signal intensity (I).   

%%%%%%%%%%%%%%%%%%%%%%%%%%%%%%%%%%%%%%%%%%%%%%%%%%%%%%%%%%%%
\section{Benchmarking in-solution $T_1$ Spin Relaxometry}
\begin{figure*}
    \centering
    \includegraphics[]{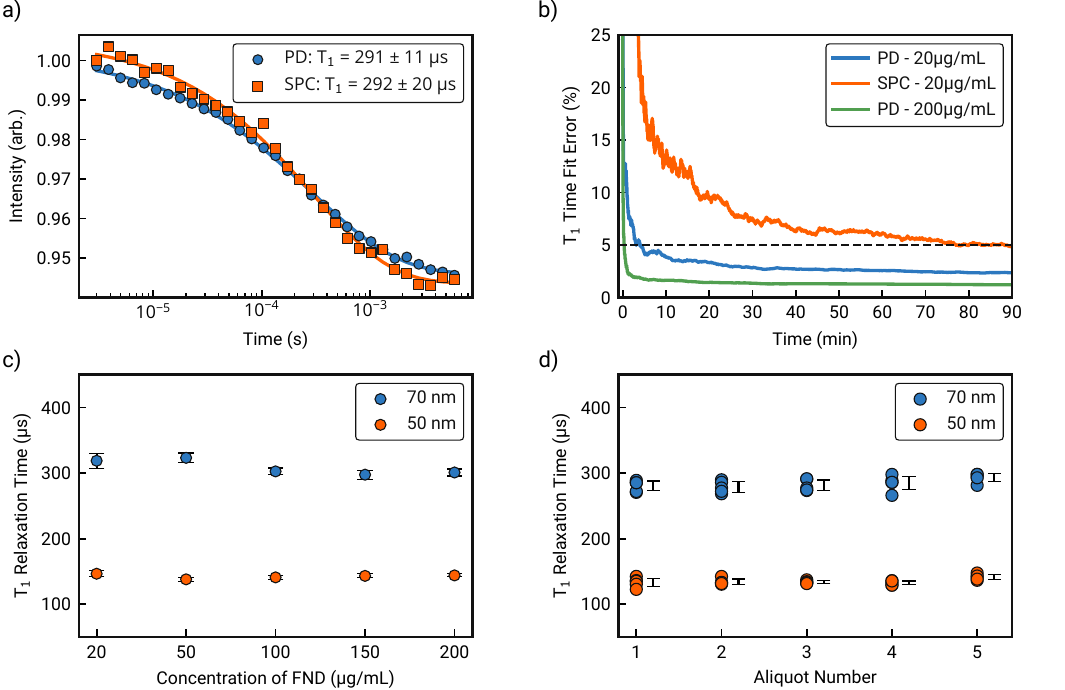}
    \caption{a)	Comparison of the $T_1$ curves obtained using the previous SPC and the new PD-based system. b) Comparison of the fit error on the $T_1$ time for the SPAD and PD systems. c) The $T_1$ relaxation time measured for different concentrations of 50 and 70~nm FNDs using the PD-based system. d) Intra- and inter-sample variation using 50 and 70~nm FNDs on the PD-based system. Five aliquots from the 50 and 70~nm samples were taken and then each measured five times. All measurements are shown.}
    \label{fig:2}
\end{figure*}

To benchmark the performance of the PD-based cuvette system for dynamic $T_1$ spin relaxation measurements, we first conducted a direct comparison with a previously reported SPAD system to evaluate signal to noise ratios (SNRs), and measurement time \cite{grantMethodInSolutionHighThroughput2023}. 
Initial $T_1$ measurements were performed using 70~nm FNDs (Adamas Nanotechnologies) dispersed in ultrapure water (Milli-Q, Millipore). 
The same solution of FNDs was measured using both approaches and the resulting $T_1$ relaxation curves are shown in Figure \ref{fig:2}a. 
The two techniques show good agreement; fitting with Equation \ref{eq:fit} yielded $T_1$ relaxation times of 290~\textpm~10 \textmu s from the PD-based system compared to 290~\textpm~20~\textmu s using the SPAD system. In addition, the spin contrast and stretch exponential power were also consistent, as shown in the Supplementary Information (SI), SI Table \ref{SI_table_fitting_values}.

To characterize the improvement in SNR, we evaluate the fit error from the $T_1$ curve as a function of time. Figure \ref{fig:2}b shows the error in the $T_1$ relaxation time from the nonlinear least squares fit of Equation \ref{eq:fit} to the data over a 90 minute period.
The fit error from the SPAD system does not plateau until the end of this 90 minute window, at a fit error of 5\%. 
Under the same conditions, the PD based system reaches the same fitting error in under 10 minutes.
Due to the saturation level of the SPAD detector ($\sim$1~Mcounts/s) the measurement speed cannot be improved by increasing the concentration of FNDs. 
The PD detector, however, has a much higher saturation limit and therefore the FND concentration can be increased, resulting in further improvements in measurement speed. Figure \ref{fig:2}c shows the measured $T_1$ relaxation times from 70 nm and 50 nm FNDs at concentrations ranging from 20~\textmu g/mL up to 200~\textmu g/mL. The $T_1$ relaxation times were consistent across this concentration range. At high FND concentrations of 200~\textmu g/mL the PD based system is able to measure a $T_1$ relaxation curve with a 5\% fit error in under 60 seconds. However, to conserve FND material, the following sensing experiments were performed with a FND concentration of 100~\textmu g/mL.  

The ability to measure a $T_1$ curve from an ensemble of FNDs within a minute is a major step towards time-resolved monitoring of chemical reactions. We note the addition of apertures in the detection path significantly reduces transmission of high angle scattered laser light, enabling us to operate at FND concentrations at least an order of magnitude higher than previously reported.

Finally, we looked at the intra- and inter-measurement variability of the 50 and 70~nm FNDs. This was achieved by taking five successive measurements from five aliquots of each FND size (Figure \ref{fig:2}d and SI Table \ref{SI_variation_table}). It is clear that the intra-measurement variability (a function of the material) is greater than the inter-measurement variability (a function of the measurement apparatus). This suggests that while the measurement methodology is sound - including the sample preparation technique - the inhomogeneity of the FNDs themselves, contributes to the overall variability between measurements. This is not surprising given the large size and shape distribution known to be present in commercial FND samples\cite{eldemrdashFluorescentHPHTNanodiamonds2023}. It is worth noting that more homogeneous distributions in particle shape and size can been achieved using etching techniques\cite{zhangAnchoredNotInternalized2017} and material synthesis approaches\cite{baoQuantumGradeNanodiamondsSingleStep2026}. 

%%%%%%%%%%%%%%%%%%%%%%%%%%%%%%%%%%%%%%%%%%%%%%%%%%%%%%
\section{Real-time monitoring of a chemical reaction}
\begin{figure*}
    \centering
    \includegraphics[]{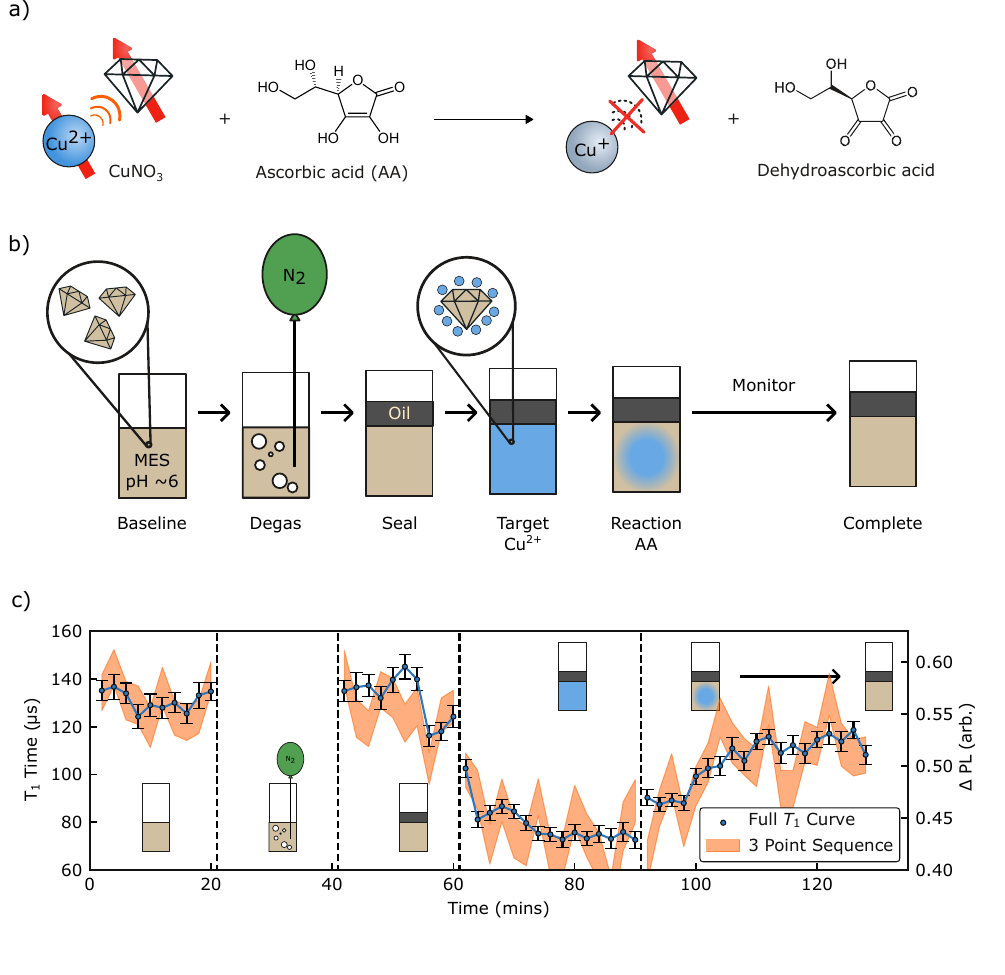}
    \caption{a) Reaction schematic of the reduction of paramagnetic Cu(II) by ascorbic acid to magnetically silent Cu(I). b) Diagram of the sample preparation and chemical reaction procedure. 50~nm FND (100~\textmu g/mL) were prepared in MES buffer (25~mM, pH 6) and degassed with nitrogen before sealing with a layer of mineral oil. An aliquot of Cu(II) solution (1~\textmu L,4~mM) was injected through the oil layer. To start the chemical reaction an excess of ascorbic acid (1 \textmu L, 500 mM) was injected through the oil barrier. c) The fitted $T_1$ time from the full $T_1$ curve at two minute intervals at each step of the reaction protocol shown in b) (blue trace). The error bars on the points represent the fit error on the $T_1$ parameter from Equation (1). Orange shaded region shows the change in PL signal and associated variance over the course of the reaction protocol reprocessed from the full trace data using a 3 point sequence. Vertical dotted lines represent each step in the protocol.}
    \label{fig:3}
\end{figure*}

Before conducting dynamic chemical sensing, we explored the measurement sensitivity of the 50 and 70~nm FNDs (calculation shown in SI). We found that although the 50~nm FNDs have a lower baseline $T_1$ time, this is offset by their comparable brightness and higher surface-area-to-volume ratio, resulting in similar sensitivities for both 50 and 70~nm FNDs. Moreover, the shorter baseline $T_1$ for the 50~nm particles allows acquisition times that are approximately half of the 70~nm FNDs, hence, the following sensing experiments were conducted with the smaller particles.

Dynamic measurements of paramagnetic reactions are demonstrated here by monitoring changes in the $T_1$ relaxation time of FND ensembles following the introduction of Cu(II) ions into the solution. Each Cu(II) ion contains a single unpaired electron spin which is a source of external magnetic field noise. The presence of this paramagnetic noise reduces the $T_1$ relaxation time of the FNDs. To illustrate the dynamic detection, we introduce a mild reductant in the form of ascorbic acid (AA). The AA performs a single electron reduction on Cu(II) ions, resulting in the generation of Cu(I) ions, which are magnetically silent, as shown in the reaction scheme in Figure \ref{fig:3}a. 
Figure \ref{fig:3}b shows each step in the measurement procedure. First, a solution of 50~nm FNDs (100~\textmu g/mL) were dispersed in 2-(N-morpholino)ethanesulfonic acid (MES) buffer (25~mM), adjusted to pH 6.
The pH was kept below 7 to limit the formation of copper hydroxide species that form under basic conditions. These species are insoluble and as such they would gradually fall out of solution, resulting in an effective decrease in target concentration.
Following this, the solution was degassed with nitrogen for 10 minutes to remove dissolved oxygen from the solution. Dissolved oxygen can react with both AA and Cu(I), making it difficult to interpret the changes in the $T_1$ relaxation time due to Cu(II) concentration alone. After purging with nitrogen, a layer of mineral oil is added to seal the solution from atmospheric oxygen\cite{dekkerOxidationAscorbicAcid1940}.
An aliquot of copper nitrate solution (1~\textmu L,4~mM) was then delivered to the FND solution, via a pipette, through the mineral oil barrier, resulting in a final Cu(II) concentration of 10~\textmu M in the solution. 
After the injection of copper nitrate, a small aliquot of concentrated AA stock solution (1 \textmu L, 500 mM) was injected into the cuvette to start the reduction reaction. 

Figure \ref{fig:3}c illustrates the $T_1$ times acquired at 2 minute intervals throughout the measurement procedure. The degassing and sealing step are shown to have no effect on the basleine $T_1$ of the FNDs.
Following the injection of the Cu(II) solution we observe a rapid decrease in $T_1$ time from the baseline of 131$\pm$5~\textmu s to 76$\pm$4~\textmu s. 
The reduction in $T_1$ time continues over several minutes as the positively charged Cu(II) ions are attracted to the negatively charged carboxyl groups on the surface of the FNDs. These groups are known to strongly complex positively charged ions\cite{manceauNatureCuBonding2010}.

%This binding of Cu(II) to the surface may explain the improved sensitivity of these FNDs compared to bulk diamond as they have a much larger surface-area-to-volume-ratio meaning comparably more of the NV defects will be in proximity to the Cu(II) ions \cite{simpsonElectronParamagneticResonance2017a}.

Once the $T_1$ time had saturated, the AA reducant (1 \textmu L, 500 mM) was added to reduce the Cu(II) ions. As expected, the $T_1$ time increased over a 20 minute period, indicating the successful reduction of Cu(II) to Cu(I). The temporal resolution of the $T_1$ measurements presented in Figure 3 was limited to 2 minutes due to the acquisition of the full $T_1$ relaxation curve.
To enhance the temporal resolution, a three-point $T_1$ sequence can be applied to infer the relative changes in the $T_1$ relaxation time. By reprocessing the data from Figure \ref{fig:3}c using only the first, last and $\frac{T_1}{2}$ $\tau$ pulse, we show that the trends in the data are maintained using this measurement sequence. 
A three-point $T_1$ sequence was chosen over a single point measurement, to mitigate against potential changes to the NV spin contrast due to the presence of charged paramagnetic species. A change in the spin contrast would lead to a perceived change in $T_1$ relaxation time if a single point at $\frac{T_1}{2}$ $\tau$ was used. 
Reducing the measurement sequence to only 3 pulses with the current implementation improved the total acquisition time by 2.3x. This is lower than the 9x improvement we expect from the reduction in measurement sequence length. The lower than expected improvement is a result of the slow string conversion and transfer from the FPGA, which can be overcome by sending the raw data directly over a TCP connection as opposed to the SCPI server ultilised in this work. These updates would allow us to reach a maximum time resolution of 15 seconds. 

Further improvements to the other areas of the system such as integration of additional collection optics such as a parabolic reflector and/or and integrating sphere could provide PL enhancements of up to 10-20x. Additionally, FNDs samples with higher NV concentrations (8-12ppm) and large area avalanche photodiodes (4 mm$^{2}$) with MHz measurement bandwidth could provide further improvements(x48) in temporal resolution, which would allow for a three point $T_1$ sequence to be obtained in less than 500~ms.

\section*{Conclusion}
Here, we have shown an in-solution $T_1$ relaxometry method that can be used for dynamic paramagnetic chemical reaction monitoring in solution. Our method achieves this result by integrating an avalanche photodiode operating in linear mode and an FPGA for high speed data acquisition. In addition, a custom 3D-printed cuvette holder was designed to accommodate the photodiode and reduce unwanted high-angle scattering of green excitation light.
This custom design increased the range of FND concentrations that could be measured; from 20~\textmu g/mL to at least 200~\textmu g/mL and reduced the integration times down to 15 sec, which is an order of magnitude faster than previous work \cite{grantMethodInSolutionHighThroughput2023}. 
We demonstrate this method of $T_1$ sensing to monitor the reduction of Cu(II) by ascorbic acid in real-time.

Although not explored here, this fast, $T_1$ measurement method also provides a means for quickly characterising $T_1$ relaxation times from FNDs. 
Efficient assessment methodologies are a boon for material development of FNDs as many of the standard characterisation techniques are laborious and require large quantities of FND powder.
Further improvements in measurement speed could be achieved with advancements in collection efficiency and improved FND material properties. By implementing a three-point measurement sequence as shown in this work, sub-second time resolution become possible, opening the pathway for monitoring the production of paramagnetic molecules such as free radicals.
Finally, by incorporating cheap of the shelf components (FPGA and avalanche photodiode) the total cost of the system has reduced by an order of magnitude, making this method highly accessible to end users. 

\section*{Acknowledgements}
This work was financially supported by the Australian Research Council (ARC)
Centre of Excellence for Quantum Biotechnology (CE230100021). D.A.S. acknowledges support from the National Intelligence and Security Discovery Research Grants Program (NS220100071) and the ARC Mid-Career Industry Fellowship (IM240100073). T.R would like to acknowledge support from the Commonwealth through an Australian Government Research Training Program Scholarship [DOI: https://doi.org/10.82133/C42F-K220] and funding from the CSIRO Industry PhD program.

\bibliography{CuvettePaper}% Produces the bibliography via BibTeX.

\end{document}

\preprint{APS/123-QED}

    \title{Method for real-time monitoring of paramagnetic reactions using spin relaxometry with fluorescent nanodiamonds}% Force line breaks with \\

    \author{Trent Ralph}
    \email{trent.ralph@unimelb.edu.au}
    \affiliation{School of Physics, The University of Melbourne}%
    \affiliation{Manufacturing, CSIRO}%
    
    \author{Lianne Lay}
    \affiliation{School of Physics, The University of Melbourne}%
    
    \author{Sepehr Ahmadi}
    \email{sepehr.ahmadi@csiro.au}
    \affiliation{Manufacturing, CSIRO}%

    \author{Erin S. Grant}
    \email{erin.grant@rmit.au}
    \affiliation{School of Science, RMIT University}%

    \author{David A. Simpson}%
    \email{simd@unimelb.edu.au}
    \affiliation{School of Physics, The University of Melbourne}%

\maketitle
\newpage
\section*{Derivation of $T_1$ sensitivity equation}
The derivation a signal-to-noise ratio for $T_1$ relaxometry is adapted from Wood et al. \cite{woodWidebandNanoscaleMagnetic2016}, and Tetienne et al. \cite{tetienneSpinRelaxometrySingle2013a}.
We first consider the NV defect ground state as a closed three-level system composed of the three spin sublevels m$_{s}$ = 0, $\pm$1 with populations n$_{0}$, $\pm$1. The spin sublevels m$_{s}$ = 0
and m$_{s}$ = $\pm$1 are coupled by two-way transition rates of strength $k_{01}$. 
\begin{align}
    n_0(\tau) & = (k_{01})\left[-\frac{\partial}{\partial \tau}n_0(\tau) + 2\frac{\partial}{\partial \tau} n_{\pm 1}(\tau)\right] \\
    n_{+1}(\tau) & =(\Gamma_{1})\left[\frac{\partial}{\partial \tau}n_0(\tau) - 2\frac{\partial}{\partial \tau} n_{+1}(\tau) + 2\frac{\partial}{\partial \tau} n_{-1}(\tau)\right] \\
    n_{-1}(\tau) & =(\Gamma_{1})\left[\frac{\partial}{\partial \tau}n_0(\tau) + 2\frac{\partial}{\partial \tau} n_{+1}(\tau) - 2\frac{\partial}{\partial \tau} n_{-1}(\tau)\right] \\
\end{align}
After initialization into the m$_{s}$ = 0 spin sublevel with an optical pulse, and solving the rate equations using the stipulation that $n_0(\tau)+2n_{\pm 1}(\tau) = 1$, gives:
\begin{align}
    n_0(\tau) &= \frac{1}{3}+\left[n_{0}(0) -\frac{1}{3}\right]\exp(-\Gamma_{1}\tau)\\
    n_{\pm 1}(\tau) &= \frac{1}{3}\pm\left[\frac{n_{+1}(0)-n_{-1}(0)}{2}\right]\exp{(-3\Gamma_{1}\tau)}-\frac{1}{2}\left[n_0(0)-\frac{1}{3} \right]\exp{(-\Gamma_{1}\tau)}\\
\end{align}
where $\Gamma_1=3k_{01}$ and in all cases $\Gamma_{1} = \Gamma_{bsl}+\Gamma_{env}$, where $\Gamma_{bsl}$ is the baseline rate governed by processes intrinsic to the diamond and $\Gamma_{env}$ is determined by the external environment. The PL intensity is given by: 
\begin{equation}
    \mathcal{I}(\tau) = A_0n_0(\tau)+A_1\left[n_{+1}(\tau)+n_{-1}(\tau) \right]
\end{equation}
where $A_0$ and $A_1$ are the PL rates associated with the spin states $m_s=0$ and $m_s=\pm1$ respectively. 
\\\\ Owing to spin-dependent PL of the NV defect, we consider $A_1$ $<$ $A_0$. Using Eq.(1), the signal $\mathcal{I}(\tau)$
can then be written as:
\begin{align}
    \mathcal{I}(\tau)=\mathcal{I}(\infty)\left[1+\mathcal{C}_1e^{-3\Gamma_{1}\tau} \right]
\end{align}
where:
\begin{align}
    \mathcal{I}(\infty) &= \frac{A_0+2A_1}{3}\\
    \mathcal{C}_1 &= \frac{A_0-A_1}{A_0+2A_1}\left[3n_0(0) -1 \right]
\end{align}
The total number of photons detected is then:
\begin{equation}
    \mathcal{N}(\Gamma_{1}, \tau) = \frac{\mathcal{R}T_{tot}t_{RO}}{t_{RO}+\tau}\left[1-\mathcal{C}_1+\mathcal{C}_1e^{-3\Gamma_{1}\tau} \right],
\end{equation}
where $R$ is the photon count rate, $T_{tot}$ is the total measurement time, and $t_{RO}$ is the readout time. If we write this in terms of the environment, $\Gamma_{env}$ and $\Gamma_{bsl}$ values, this becomes:
\begin{align}
    \mathcal{N}(\Gamma_{1}, \tau) = \frac{\mathcal{R}T_{tot}t_{RO}}{t_{RO}+\tau}\left[1-\mathcal{C}_1+\mathcal{C}_1e^{-3(\Gamma_{bsl}+\Gamma_{env})\tau} \right]
\end{align}
At a specific evolution time, $\tau$, the difference in the number of photons caused by spins in the environment, compared to the baseline value, is given by:
\begin{align}
    \Delta\mathcal{N}_{signal}(\tau) &= \mathcal{N}(o,\tau) - \mathcal{N}(\Gamma_{env}, \tau)\\
    &= \frac{\mathcal{R}T_{tot}t_{RO}}{t_{RO}+\tau}\left[\mathcal{C}_1e^{-3(\Gamma_{bsl})\tau} -\mathcal{C}_1e^{-3(\Gamma_{bsl}+\Gamma_{env})\tau} \right]\\
    &= \frac{\mathcal{R}T_{tot}t_{RO}}{t_{RO}+\tau}\mathcal{C}_1e^{-3(\Gamma_{bsl})\tau}\left[1 - e^{-3\Gamma_{env}\tau} \right]\\
\end{align}
The shot noise associated with the measurement is:
\begin{align}
    \Delta \mathcal{N}_{noise}(\tau) &= \sqrt{\mathcal{N}(\Gamma_{1,},\tau)}\\
    &\approx \sqrt{\frac{\mathcal{R}T_{tot}t_{RO}}{t_{RO}+\tau}}
\end{align}
assuming that $\mathcal{C}<<1$. The signal to noise, is then:
\begin{align}
    SNR(\tau) &= \frac{\Delta\mathcal{N}_{signal}(\tau)}{\Delta\mathcal{N}_{noise}(\tau)}\\
    &\approx \sqrt{\frac{\mathcal{R}T_{tot}t_{RO}}{t_{RO}+\tau}}\mathcal{C}_1e^{-3(\Gamma_{bsl})\tau}\left[1 - e^{-3\Gamma_{env}\tau} \right]\\
    \label{eq:SNR}
\end{align}
Applying the measured parameters from the 50 and 70~nm FNDs into Eq. \ref{eq:SNR} leads to similar SNR for both particle sizes withthe 70~nm FNDs $<4\%$ more sensitive than the 50~nm particles).

\section{Fitting parameters from the in solution cuvette methods}
As described in the main text, control $T_1$ measurements where performed using the SPC and PD cuvette system. The Table below provides comparative $T_1$ times, spin contrat and the stretched exponential power obtained from both systems.

\begin{table*}[h]
\setlength{\tabcolsep}{10pt}
\caption{Summary of the fitted values for the comparison measurement conducted on the SPC and PD systems}
\begin{tabular}{lllll}
\toprule
System & T1 relaxation time (\textmu s) & Stretch power & contrast (\%) &  \\ \midrule
PD     &      \multicolumn{1}{c}{ $291 \pm 11$ }  &  \multicolumn{1}{c}{$0.542 \pm 0.018$}  &  \multicolumn{1}{c}{$5.39 \pm 0.05 $}        &  \\
SPC    &          \multicolumn{1}{c}{ $292 \pm 20$}           &  \multicolumn{1}{c}{$0.598 \pm 0.039 $} & \multicolumn{1}{c}{$5.63 \pm 0.09 $}         &  \\
       &                         &         &          & 
\end{tabular}
\label{SI_table_fitting_values}
\end{table*}

It can be seen from the results in Table 2 that the intra-measurement variability (a function of the material) is greater than the inter-measurement variability (a function of the measurement apparatus).

\begin{table*}[h]
\setlength{\tabcolsep}{6pt}
\caption{Inter and intra-measurement variation from 50 and 70~nm FNDs.}
\begin{tabular}{@{}llllllll@{}}
\toprule
Sample & \begin{tabular}[c]{@{}l@{}}Aliquot\\ 1 (\textmu s)\end{tabular} & \begin{tabular}[c]{@{}l@{}}Aliquot \\ 2 (\textmu s)\end{tabular} & \begin{tabular}[c]{@{}l@{}}Aliquot\\ 3 (\textmu s)\end{tabular} & \begin{tabular}[c]{@{}l@{}}Aliquot\\ 4 (\textmu s)\end{tabular} & \begin{tabular}[c]{@{}l@{}}Aliquot\\ 5 (\textmu s)\end{tabular} & \begin{tabular}[c]{@{}l@{}}Average \\ (\textmu s)\end{tabular} & \begin{tabular}[c]{@{}l@{}}Standard\\ deviation (\textmu s)\end{tabular} \\ \midrule
50 nm  &  $133\pm 4$ & $134\pm 3$& $134\pm 3$ & $132\pm 3$ & $141\pm 3$ & 135 &  3   \\
70 nm & $280\pm 6$ & $279\pm 6$ & $281\pm 6$ & $285\pm 7$ & $293\pm 6$ & 284 & 5    \end{tabular}
\label{SI_variation_table}
\end{table*}
\\

\section{Control with Nitric Acid}
\begin{figure}[h]
    \centering
    \includegraphics[width=12cm]{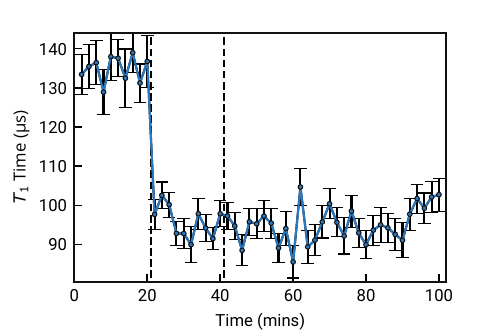}
    \caption{Plot of control experiment with 50nm FND (100~\textmu g/mL) showing no recovery of $T_1$ relaxation time after addition of $\mathrm{HNO_3}$. After a baseline was recorded an as aliquot of $\mathrm{Cu(NO_3)_2}$ (0.6~\textmu L,4~mM) was added resulting in a decrease in $T_1$ relaxation time. After stabilising an aliquot of $\mathrm{HNO_3}$ (2.5~\textmu L,100~mM) was added resulting in no change to the $T_1$ relaxation time. Vertical lines represent where each addition was performed and the error bars represent the fit error of the $T_1$ value.}
    \label{SI_fig:1}
\end{figure}

To ensure that a pH change was not responsible for the change in $T_1$ time in the copper reduction reaction a control experiment was performed using Nitric Acid at the same concentration as the ascorbic acid (AA) described in the manuscript (Figure \ref{SI_fig:1}). 
Nitric acid is not a reducing agent and should not participate in the chemical reduction of Cu(II) ions. The control experiment begins by monitoring the T1 time of the 50 nm FNDs (100~\textmu g/mL) at 2 minute intervals in (MES, 25~mM, pH~6). An aliquot of $\mathrm{Cu(NO_3)_2}$ (1~\textmu L,4~mM) was then added to the solution. The $T_1$ time decreased due to the presence of Cu(II) ions in a manner consistent with the results reported in Figure 3c of the main text. Once stabilised ($\sim$ 20 mins) an aliquot of $\mathrm{HNO_3}$ (2.5~\textmu L,100~mM) was added to give a final acid concentration of 1~mM. No recovery of the $T_1$ relaxation time was observed after the addition of nitric acid confirming that the ascorbic acid is responsible for the observed changes in $T_1$ in Figure 3 of the main text. 

\bibliography{CuvettePaper}% Produces the bibliography via BibTeX.